\documentclass[lettersize,journal]{IEEEtran}
\usepackage{amsmath,amsfonts}
\usepackage{algorithmic}
\usepackage{algorithm}
\usepackage{array}
\usepackage[caption=false,font=normalsize,labelfont=sf,textfont=sf]{subfig}
\usepackage{textcomp}
\usepackage{stfloats}
\usepackage{booktabs}
\usepackage{url}
\usepackage{multirow}
\usepackage{verbatim}
\usepackage{graphicx}
\usepackage{cite}
\usepackage{xcolor}
\usepackage{color, colortbl}
\newcommand{\ci}[1]{\tiny{±#1}}
\definecolor{Gray}{gray}{0.94}

\begin{document}

\title{An overview of neural architectures for self-supervised audio representation learning from masked spectrograms}

\author{Sarthak Yadav\IEEEauthorrefmark{1}\IEEEauthorrefmark{2}, Sergios Theodoridis\IEEEauthorrefmark{1}\IEEEauthorrefmark{3}~\IEEEmembership{Life Fellow,~IEEE} and Zheng-Hua Tan\IEEEauthorrefmark{1}\IEEEauthorrefmark{2}~\IEEEmembership{Senior Member,~IEEE}

\thanks{\IEEEauthorrefmark{1}Department of Electronic Systems, Aalborg University, Aalborg, Denmark\\\IEEEauthorrefmark{2}Pioneer Centre for Artificial Intelligence, Denmark \\\IEEEauthorrefmark{3}National and Kapodistrian University of Athens, Athens, Greece}}



\maketitle

\begin{abstract}
In recent years, \textit{self-supervised} learning has amassed significant interest for training deep neural representations without labeled data. One such self-supervised learning approach is \textit{masked spectrogram modeling}, where the objective is to learn semantically rich contextual representations by predicting removed or hidden portions of the input audio spectrogram. 
With the \textit{Transformer} neural architecture at its core, masked spectrogram modeling has emerged as the prominent approach for learning general purpose audio representations, a.k.a. audio foundation models. 
Meanwhile, addressing the issues of the Transformer architecture, in particular the underlying \textit{Scaled Dot-product Attention} operation, which scales quadratically with input sequence length, has led to renewed interest in recurrent sequence modeling approaches. Among them, \textit{Selective structured state space models} (such as Mamba) and \textit{extended Long Short-Term Memory} (xLSTM) are the two most promising approaches which have experienced widespread adoption.
While the body of work on these two topics continues to grow, there is currently a lack of an adequate overview encompassing the intersection of these topics. In this paper, we present a comprehensive overview of the aforementioned research domains, covering masked spectrogram modeling and the previously mentioned neural sequence modeling architectures, Mamba and xLSTM. 
Further, we compare Transformers, Mamba and xLSTM based masked spectrogram models in a unified, reproducible framework on ten diverse downstream audio classification tasks, which will help interested readers to make informed decisions regarding suitability of the evaluated approaches to adjacent applications.
\end{abstract}


\section{Introduction}
\IEEEPARstart{I}{n} recent years, deep representation learning has observed an immense increase in scale, both in terms of parametric complexity of deep neural networks (DNNs) as well as the amount of data required to train them. 
As such, going beyond the realm of supervised learning on labeled data and leveraging the vast amounts of unlabeled data available at our disposal has garnered significant interest. 
While several such methods exist, self-supervised learning (SSL) has emerged as a prominent approach for training deep neural representations without labeled data. Lying somewhere in the middle of the supervised learning-unsupervised learning spectrum, the training objective in SSL is to solve a \textit{pretext} supervised learning task utilizing labels automatically generated from a subset of the data itself. The sole purpose of these designed pretext tasks is to guide the learning of contextually and semantically rich representations that are useful for subsequent practical \textit{downstream} applications.
While several methods for learning self-supervised representations have been proposed, predicting removed/hidden portions of input data, a.k.a. \textit{masked predictive modeling}, which was originally proposed for natural language processing (NLP) \cite{bert2019}, is one of the most popular SSL approaches.
Together with the Transformer architecture \cite{vaswani2017attention}, 
masked predictive models have enabled several key breakthroughs across multiple application domains, such as NLP \cite{bert2019, Lan2020ALBERT, conneau-etal-2020-unsupervised,raffel2020exploring}, computer vision \cite{xie2022simmim, he2022masked, bao2022beit, tong2022videomae}, as well as speech \cite{baevski2020wav2vec, hsu2021hubert, chung21w2vbert, chen2022wavlm, pmlr-v162-baevski22, data2vec2} and audio \cite{gong2022ssast, huang2022masked, niizumi2022masked, yadav2024masked} representations. 


Although a very large proportion of recent works focus on learning self-supervised speech representations, learning general audio representations, viz. audio foundational models, has recently garnered a lot of interest. As opposed to the predominantly local information modelled by speech representations, audio foundation models need to capture a mix of local and global dependencies in the data.
Audio foundation models are often pretrained on large audio datasets to learn a general representation that can capture a wide variety of acoustic and semantic information and can be adapted to a specific downstream task without significant training efforts. 
These downstream tasks can span linguistic, paralinguistic, music/pitch perception and general audio classification domains. 
Masked predictive modeling on audio spectrogram patches, which we hereafter refer to as masked spectrogram modeling, has emerged as a prominent approach for learning general audio representations. 
In masked spectrogram modeling, input audio spectrograms are divided into (generally) non-overlapping patches. Similar to MLMs, masked spectrogram models (MSMs) comprise of a Transformer encoder that captures contextual representations from the partially masked input patches, and a decoder that reconstructs masked portions of the input based on the encoded representations. After pretraining, the decoder is usually discarded.
Based on how input patches are masked and presented to the encoder, MSMs can be broadly categorized into two types: (i) the SSAST \cite{gong2022ssast,yadav2024audiomambaselectivestate} class of models, where patches to be hidden are replaced with a learnable mask token before being fed to the encoder; and (ii) the Masked Autoencoder (MAE) \cite{he2022masked,huang2022masked,baade_mae-ast_2022, dinkel24b_interspeech} class of models, where patches to be hidden are \textit{removed} from the input sequence fed to the encoder. Both of these architectures have distinct characteristics and trade-offs. SSAST based models are simpler and closer to the original MLMs, and since the encoder is masking-aware, a computationally cheaper decoder can be employed. On the other hand, the MAE design allows for asymmetrically large encoders to be paired with much smaller decoders, as the larger computation load for the encoder is offset by smaller input sequence lengths when pretraining, making them highly scalable. 
However, while the masking methodology forces MAE encoders to learn better contextual representations, the MAE encoder design is closely coupled with the Transformer architecture and the underlying SDPA operation that treats input as elements of a set, which makes it possible to drop elements in the input sequence. This masking and encoding methodology is incompatible with traditional sequence modeling paradigms.

While Transformers have emerged as the eminent neural architecture and are the backbone of all of the above-mentioned MSMs, they are not without drawbacks. Scaled dot-product attention, the operation at the heart of the Transformer, has quadratic complexity with respect to input sequence length, and thus scales very poorly to large sequences. Further, standard scaled dot-product attention also necessitates storing the entire key-value (KV) cache for retrieval operations, thus imposing high memory requirements. Significant research efforts have been made to address these shortcomings, such as finding linear approximations for self-attention \cite{katharopoulos2020Transformers, wang2020linformer}, optimized hardware-aware implementations \cite{dao2022flashattention, dao2024flashattention} and bypassing attention entirely, either through new token mixing techniques \cite{alberti2023sumformer, mai_hypermixer_2023}, or by revisiting the classical recurrent neural network paradigm. Recently, two such approaches have garnered significant research interest: state space models (SSMs) \cite{gu2022efficiently}, and the extended long-short term memory (xLSTM) neural architecture \cite{beck2024xlstmextendedlongshortterm}. SSMs are a class of sequence models at the intersection of convolutional neural networks, recurrent neural networks and classical state spaces from control theory, and are governed by a set of first-order differential equations. While several SSM variants \cite{nguyen2022snd, fu2023hungry, poli2023hyena, sun2023retentive, peng2023rwkv} have been proposed, the most prominent is selective structured state spaces, a.k.a. Mamba \cite{gu2023mamba, mambav2}, which has demonstrated competitive performance against Transformer based alternatives in several domains, and have also seen widespread application in the audio domain \cite{speechmamba, li2024spmamba, shams2024ssamba, hamza24aum, mu2024seld, yadav2024audiomambaselectivestate}. The xLSTM architecture, as the name suggests, is an extension of the LSTM \cite{hochreiter1997long} sequence modeling approach. xLSTMs address the critical weak points in LSTMs by integrating recent technological advances resulting from years of Transformer and large language modeling research, and have demonstrated superior sequence length extrapolation capabilities compared to Transformers. xLSTMs have seen widespread adoption in several domains, including audio and speech \cite{yadav2025axlstms, kuhne2025xlstm}. Both Mamba and xLSTMs have been used for training audio foundation models in a masked spectrogram modeling paradigm.

Despite the large interest in masked spectrogram modeling and neural sequence modeling architectures beyond Transformers, a comprehensive overview article that investigates MSMs while encorporating a study of Transformers, Mamba and xLSTMs is still lacking. This paper presents a systematic overview of the aforementioned research domains. The objective of this paper is to enable the reader to navigate the scientific landscape as well as to get a good grasp of the fundamental concepts at the intersection of these topics, and hopefully inspire new lines of research in the field. We provide a comprehensive, systematic description of masked spectrogram modeling, both SSAST and the MAE class of MSMs and the various underlying components. We also provide a comprehensive coverage and review of the Mamba and the xLSTM neural sequence modeling architectures.
Further, we provide a comprehensive, consistent and reproducible empirical evaluation of Transformer, Mamba and xLSTM within an SSAST-style masked spectrogram modeling framework on a suite of 10 varied utterance level audio classification tasks, based on our previous research \cite{yadav2024audiomambaselectivestate, yadav2025axlstms}, allowing the reader to evaluate which sequence modeling approach best fits their needs. In addition to our previous work, we also evaluate how the covered sequence modeling approaches fare with different input sequence lengths and input audio clip durations.

Throughout the paper, we also discuss current challenges and possible future research directions. 
We also provide direct links to implementations and demos wherever possible, as well as source code(s) and pretrained models that can be used to reproduce the results presented in this study.

The rest of the paper is organised as follows. Section ~\ref{sec:msm} covers MSMs and relevant concepts and components. Section ~\ref{sec:seqmods} provides a comprehensive overview of Mamba and xLSTM, the two neural sequence modeling architectures in spotlight. Section ~\ref{sec:approach} covers the SSAST-style masked spectrogram modeling framework under which we investigate Transformer, Mamba and xLSTM based MSMs, as well as commentary on datasets and metrics used for the study. This is followed by Section \ref{sec:emp} where we discuss the outcomes of the empirical evaluation, along with select ablation studies. Section \ref{sec:closing} discusses the things that we could not cover in the presented analysis. Finally, Section \ref{sec:conclusion} presents our concluding remarks.

\section{Transformer based Masked spectrogram modeling}
\label{sec:msm}

\begin{figure}[!t]
    \centering
    \includegraphics[width=\linewidth]{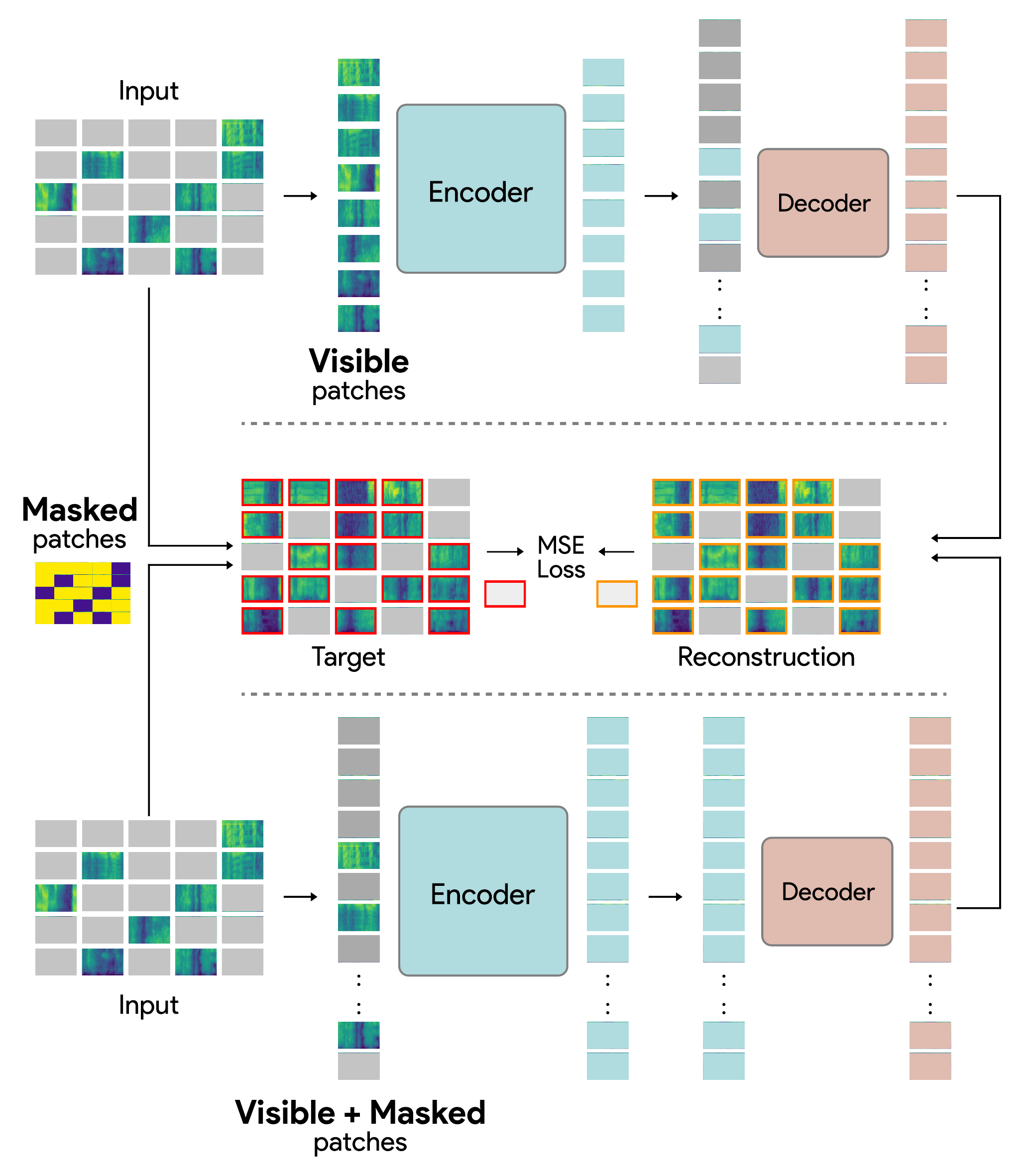}
    \caption{The main types of masked spectrogram models: MAE-style (top), where the encoder operates only on the visible patches, and SSAST-style (bottom), where masked patches are replaced by learnable mask tokens before feeding to the encoder.}
    \label{fig:msm}
\end{figure}

Masked language models such as BERT \cite{bert2019} learn strong semantic language representations without supervision by learning to predict masked portions of tokenized input text using visible, unmasked portions. Masked spectrogram models (MSMs) operate on the same underlying principles. The input audio spectrogram is divided into (generally) non-overlapping patches, and a portion of these patches is then randomly masked. MSMs learn contextualized audio representations by forcing them to reconstruct masked out patches based on the visible ones. Upcoming is a comprehensive discussion of the core components in masked spectrogram modeling, along with commentary on how these components differ across SSAST and MAE style MSMs, wherever applicable. An overview of MSMs can be found in Figure~\ref{fig:msm}.
\subsection{Creating and embedding audio spectrogram patches}
\label{ssec:patching}

The first step is to create non-overlapping audio spectrogram patches from an input audio spectrogram. The reasoning behind computing non-overlapping patches is to avoid information leakage across different patches, which would make the reconstruction task easier and result in worse performance, as demonstrated by \cite{huang2022masked}.

Let $\mathbf{X} \in \mathbb{R}^{\mathtt{T} \times \mathtt{F}}$ denote the input audio spectrogram with $\mathtt{T}$ frames and $\mathtt{F}$ frequency bins. Then, we extract $N$ non-overlapping patches of shape $t \times f$ along the time and frequency axes, respectively, and flatten them to yield patched input spectrograms $\mathbf{X}_p \in \mathbb{R}^{N \times t.f}$. Thus, while $\mathbf{X}_p$ is a patched representation, it is worth noting that the representation space has not changed: they are essentially rearranged spectrograms. 
A learnable linear projection embeds them to a $d_{\text{enc}}$ dimensional space, which yields patched input $\mathbf{X}_e \in \mathbb{R}^{N \times d_{\text{enc}}}$ ready for futher processing. We can now optionally append a representative learnable class token $\mathtt{cls} \in \mathbb{R}^{d_{\text{enc}}}$ to the beginning of $\mathbf{X}_e$. 
While not essential, the $\mathtt{cls}$ token acts like a \textit{register} for the Transformer blocks to store global information in \cite{darcet2024vision} and facilitate straightforward downstream fine-tuning of the MSM. Register tokens have also been shown to enable interpretable attention maps in Transformers \cite{darcet2024vision}.
For the sake of clarity and consistency, we omit the $\mathtt{cls}$ token in any equations and notation.
Finally, to encode positional information, positional embeddings are added to $\mathbf{X}_e$. The vast majority of MSMs use fixed sinusoidal positional embeddings \cite{gong2022ssast,huang2022masked,niizumi2022masked,yadav2024masked,yadav2024audiomambaselectivestate,yadav2025axlstms}. However, if a different input duration is expected during test-time, appropriate measures, such as positional embedding extrapolation, or a better suited positional embedding, like Rotary Positional Embeddings (RoPE) \cite{su2024roformer}, can be used. 
    
\subsubsection*{Impact of patch size}

The shape of the patch, $t \times f$, is an important hyperparameter and needs to be optimized together with the spectrogram computation parameters. Selection of $t$ and $f$ dictates the time-frequency resolution of the resulting patches and, effectively, the entire MSM, and the smaller the value, the finer the resolution. This allows MSMs to implicitly model both temporal and frequency structure from the input spectrogram, in contrast to several masked predictive modeling based speech representations, such as \cite{liu2020mockingjay,baevski2020wav2vec,hsu2021hubert,liu2021tera,chen2022wavlm} where masking is applied only along the temporal axis. Let us consider some common cases:
\begin{enumerate}
    \item $f < \mathtt{F}$, $t < \mathtt{T}$, $f = t$: A square patch, which is quite prevalent and is adopted by a large number of approaches \cite{gong2022ssast,huang2022masked}. For a given spectrogram, this results in a neutral setting without trading off time or frequency resolution for the other.
    \item $f < \mathtt{F}$, $t < \mathtt{T}$, $f \neq t$: Several works have evaluated rectangular patches that offer different time-frequency resolution tradeoffs \cite{niizumi2022masked,yadav2024masked, yadav2024audiomambaselectivestate, yadav2025axlstms}. Most notably, \cite{niizumi2022masked} conducted in-depth analysis with different patch resolutions, and demonstrate that patches with finer temporal resolutions performed better for speech based tasks.
    \item $f = \mathtt{F}$, $t = 1$: In this case, there is a patch for every frame in your input spectrogram, and every patch embeds frequency information from all the frequency bins. This setting is maximizing time resolution at a tradeoff for frequency resolution, and has been shown to work well for speech based applications, such as Keyword Spotting \cite{berg2021keyword, holgerkwt1, holgerkwt2}.
\end{enumerate}

\subsection{Masking strategies}
\label{ssec:maskingstrats}

As previously stated, in the same vein as MLMs like BERT \cite{bert2019}, a portion of the embedded patches ($\mathbf{X}_e$) will be masked out. 
In contrast to MLMs, where masking between $10-20\%$ of the input tokens is sufficient to learn good representations, MSMs have very high masking ratios (50-80\%) similar to their image counterparts \cite{bao2022beit, he2022masked}. In contrast to language, which is very information dense, spectrograms (especially high resolution spectrograms), can be highly spatially redundant. As a result, recovering missing patches is easier in the case of MSMs, and such a high portion of random masking is necessary to learn good representations. From here on, we will refer to masking ratio as $m_r \in (0,1)$, which denotes the proportion of patches to be masked.

In literature, patch masking methologies can be broadly classified into 3 categories:
\begin{enumerate}
    \item Unstructured random masking, which is truly random masking of individual patches.
    \item Structured random masking, where the mask follows a shape or structure. Examples include horizontal/vertical strips \cite{niizumi2022masked} and block-wise or grid masking, where a span of contiguous blocks or a grid are masked out.
    \item Guided masking strategies, where the objective is to find optimal masking strategies based on the input. These include masking based on semantic concepts to hide \cite{li2022semmae} as well as masking based on the emerging concepts in the model itself \cite{shin2024selfguided}, to name a few.
\end{enumerate}
Most MSMs in literature utilize an unstructured random masking strategy, where tokens are masked out completely at random. 
This strikes the best balance between computational overhead, ease of implementation as well as performance on a varied bunch of tasks. However, selecting a structured or refined masking strategy based on domain-specific information for a particular application can pay off.
Finally, finding optimal model or data informed masking strategies, such as using adversarial learning to determine which patches to mask for a given input, can be a potential avenue for future research.

\subsection{Encoding contextual representations}
\label{ssec:encoding}

After creating and embedding patches as well as computing the mask indices, it is time to feed the patches to the encoder, which is a stack of Transformer blocks of $d_{\text{enc}}$ dimensions. 
Since the encoder is operating on a partial view of the input spectrogram (patches), it needs to learn high-level semantic information from the limited view and encode it into its outputs, so that the decoder can then reconstruct the missing input from them.
The structure of the encoder is one of the key differences between SSAST-style and MAE-style MSMs. 
In SSAST-style models, the spectrogram patches to be masked, denoted by the stored mask indices, are replaced in-place by a learnable \textit{mask token}, i.e. $\mathbf{X}_e \in \mathbb{R}^{N \times d_{\text{enc}}} \mapsto \mathbf{X}_\text{enc} \in \mathbb{R}^{N \times d_{\text{enc}}}$, where $\mathbf{X}_\text{enc}$ denotes the input to the encoder. 
In contrast, in MAE-style models, the patches to be masked are \textit{removed} from the input sequence using an index operation, i.e. $\mathbf{X}_e \in \mathbb{R}^{N \times d_{\text{enc}}} \mapsto \mathbf{X}_\text{enc} \in \mathbb{R}^{(1-m_r)N \times d_{\text{enc}}}$, where $m_r$ is the masking ratio. 
As a result, the encoder in an MAE-style MSM ingests only a small fraction of the total number of input patches, allowing the usage of much larger encoders compared to an SSAST-style MSM for the same number of multiply-accumulate operations (MACs).
This is made possible by the Transformer blocks, where the underlying scaled dot-product attention operation do not treat the input as a sequence, but rather as elements in a set of inputs with the computed attention weights measuring relative importance of each element with respect to every other. Thus, the MAE encoder is closely coupled with the Transformer, and is fundamentally incompatible with other sequence modelling architectures.
Thus, the encoder returns $\mathbf{Z}_{\text{enc}} = \text{Encoder}(\mathbf{X}_e)$, where $\mathbf{Z}_{\text{enc}} \in \mathbb{R}^{N \times d_{\text{enc}}}$ for SSAST-style models and $\mathbf{Z}_{\text{enc}} \in \mathbb{R}^{(1-m_r)N \times d_{\text{enc}}}$ for MAE-style models.

The majority of the MSMs incorporate encoders comprised of standard Transformer layers with global self-attention, which might be suboptimal for audio recognition tasks where discriminative information is predominantly local. As a result, several works have evaluated MSMs with encoders based on modified Transformer blocks, such as the Swin Transformer \cite{liu_swin_2021}, which can explicitly capture local and global information \cite{huang2022masked, yadav2024masked}. This is also a noteworthy avenue for potential future research, since different audio recognition tasks operate at a different interplay of local-global dynamics. For instance, a hybrid, adaptive attention formulation might be able to give us the best of both worlds. 

\subsection{Reconstruction and loss computation}
\label{ssec:decoder}

Once the visible patches have been embedded into $\mathbf{Z}_{\text{enc}}$, the decoder's task is to reconstruct the original spectrogram patches $\mathbf{X}_p$ corresponding to the masked indices based on the high-level contextual information encoded in $\mathbf{Z_\text{enc}}$ by the encoder. 
First, we linearly project $\mathbf{Z}_{\text{enc}}$ into a $d_{\text{dec}}$ dimensional space. In MAE-style MSMs, learnable \textit{mask tokens} are then inserted in a manner that restores the correct order of the masked-unmasked positions, followed by injecting positional information again into the resulting vector using a positional embedding. 
In contrast, this operation is not needed for SSAST-style MSMs, since mask tokens were inserted prior to encoding and thus the $\mathbf{Z_\text{enc}}$ representation is complete. At this point, the number of tokens in both $\mathbf{X}_p$ and the intermediate representation $\mathbf{Y}_d \in \mathbb{R}^{N \times d_{\text{dec}}}$ is $N$. Now, we need to project $\mathbf{Y}_d$ to $t.f$ dimensions so that we can compute reconstruction loss. 

In the case of SSAST-style MSMs, this is generally done using a linear layer with $(t.f)$ output features. In contrast, in the literature MAE-style MSMs usually include a stack of Transformer blocks with $d_{\text{dec}}$-dimensions before the final linear layer projecting to $t.f$. However, the decoder Transformer does not need to be as complex as the encoder, and can often have a fraction of the layers, as shown in \cite{niizumi2022masked, yadav2024masked}. Some recent papers have even shown that there is no need for a Transformer-based decoder, and a convolutional decoder can be used instead \cite{data2vec2, gao22mcmae}. Either way, it is clear that due to their design, the decoding is more nuanced in the case of a MAE-style MSM. 

The decoder yields the final reconstructed output $\mathbf{Y'} = \text{Decoder}(\mathbf{Z}_\text{enc})$ such that $\mathbf{Y}' \in \mathbb{R}^{N \times t.f}$, which has compatible shapes with $\mathbf{X}_p$. 
The objective function for pretraining the MSM is the mean squared error (MSE) between the \textit{masked} spectrogram patches $\mathbf{X}_p$ and the corresponding reconstructions $\mathbf{Y}'$, which has shown better performance versus computing MSE for all the patches \cite{gong2022ssast, he2022masked, huang2022masked}. While most MSMs use only a reconstruction objective, some papers also adopt auxiliary loss functions. For instance, \cite{gong2022ssast, baade_mae-ast_2022} also use an InfoNCE loss function to create a dual discriminative + reconstructive objective function, while other works introduce a contrastive learning loss function as an auxiliary objective \cite{deshmukh2023pengi, gong2022ssast}. Furthermore, recent approaches like \cite{chenbeats23} completely change the target task from regression to a classification objective with self-distilled acoustic units as labels. Thus, while reconstruction alone yields good performance, looking at auxiliary loss functions that suit ideal representation characteristics can be useful, and is an interesting avenue for future research.

\section{Sequence modeling beyond Transformers}
\label{sec:seqmods}

\subsection{On Transformers and Scaled Dot-Product Attention}

Transformers have emerged as the eminent neural architecture. Starting out as a new sequence modeling paradigm for NLP applications, Transformers have seen quick and widespread adoption across different input modalities and application domains. While there is a cause-and-affect relationship between emergent phenomena across data modalities, such as the concept of a patch as the atomic input unit in computer vision and audio, the prime drivers of this adoption are the strong generalization abilities and the inherently modality agnostic design of Transformers, including the scaled dot-product attention (SDPA) operation at their core. However, Transformers also have several drawbacks.

\subsubsection*{Quadratic Complexity}
Recall that the following equation can be used to describe the self-attention operation:
\begin{equation}
    \text{Attention}(\mathbf{Q},\mathbf{K},\mathbf{V}) = \text{softmax}\left(\frac{\mathbf{Q}.\mathbf{K}^T}{\sqrt{d_k}}\right)\mathbf{V},
    \label{eq1}
\end{equation}
where $\mathbf{Q},\mathbf{K},\mathbf{V} \in \mathbb{R}^{N \times d_k}$. Here, the matrix multiplication for computing attention weights $\mathbf{Q}.\mathbf{K}^T$ requires $N \times N$ operations and storage, and thus has quadratic compute and space complexity with respect to sequence length $N$, which scales poorly for very large $N$. Furthermore, retrieval operations for a new query ($\mathbf{Q}$) requires storing the entire Key-Value (KV) cache.

\subsubsection*{Standard attention is global}

As evident from Eq~\ref{eq1}, attention computation is a global operation, as the process of attention matrix computation takes all $N$ elements of $K$ and $Q$ into consideration. While this has the benefit of computing global long-range dependencies in the input, such an operation is inefficient for tasks that comprise predominantly of local information, such as speech. 
This is evidenced by the fact that compared to the standard Transformer, models that explicitly employ techniques that excel at modeling local information, such as convolutional layers, excel in speech recognition (e.g. Conformer \cite{gulati20_interspeech}), and in other applications such as visual object detection and localization, where capturing fine-grained information is critical.

A lot of research efforts have been spent on addressing these shortcomings. 
Several papers attempt to address the quadratic complexity of Transformers by proposing sub-quadratic or linear approximations for SDPA \cite{katharopoulos2020Transformers,wang2020linformer, choromanski2021rethinking, meng2025polaformer}. 
A lot of hardware aware implementations that improve real-world performance characteristics of Transformers, such as \textit{FlashAttention} \cite{dao2022flashattention,dao2024flashattention} have also been proposed. 
Other methods, such as Multi-query attention (MQA) \cite{shazeer2019mqa} and Grouped-query attention (GQA) \cite{ainslie2023gqa} attempt to reduce the number of processed key-value pairs in order to improve runtime performance. 
Methods such as Ring Attention \cite{liu2024ringattention} attempt to address the high memory requirements of SDPA and enable scaling Transformers to very large context lengths, whereas hybrid local-global architectures offer better theoretical complexity \cite{liu_swin_2021} as well as improved local and hierarchical feature modeling \cite{liu_swin_2021,gulati20_interspeech} capabilities. Finally, several approaches attempt to completely replace standard attention with other alternatives \cite{lee-thorp_fnet_2022, alberti2023sumformer, mai_hypermixer_2023}. 

Needless to say, improving upon Transformers and SDPA is an area with incredibly high research interest. While there have been a lot of methods that attempt to address these issues, two recent independent lines of research have showed a lot of promise, namely selective structured state space models (a.k.a. Mamba) \cite{gu2023mamba, mambav2} and extended long short term memory (xLSTM) \cite{beck2024xlstmextendedlongshortterm}, both of which are recurrent neural networks. We discuss these two methods in more detail in the following sections.

\subsection{Selective Structured State Space Models}
\label{ssec:ssms}

Recently, \cite{gu2022efficiently} proposed structured state space sequence models (S4) that excel at modeling long range dependencies in very long sequences. S4 models are related to CNNs, RNNs and classical state space models, and are based on a simple continuous system (Eq.~\ref{eq:s4c}).
\begin{subequations}
\label{eq:s4c}
    \begin{align}
        h'(t) &= \mathbf{A}h(t)+\mathbf{B}x(t) \label{eq:s4c1}\\
        y(t) &= \mathbf{C}h(t)\label{eq:s4c2}
    \end{align}
\end{subequations}
S4 models are linear time-invariant (LTI) models governed by four parameters $(\Delta,\mathbf{A},\mathbf{B},\mathbf{C})$, and constitute a sequence-to-sequence transformation on input $\mathbf{x} \in \mathbb{R}^{T}$ through a latent space of dimensionality $N$ that is comprised of two stages: 
\subsubsection*{1. Discretization stage}
    which involves zero-order hold (ZOH) discretization of parameters $\mathbf{A}$ and $\mathbf{B}$ (Eq.~\ref{eq:s4disc})\begin{subequations}
    \label{eq:s4disc}
        \begin{align}
            \overline{\mathbf{A}} &= \text{exp}(\Delta \mathbf{A}) \label{eq:s4disc1}\\
            \overline{\mathbf{B}} &= (\Delta \mathbf{A})^{-1}(\text{exp}(\Delta \mathbf{A}) - \mathbf{I}).\Delta \mathbf{B}\label{eq:s4disc2}
        \end{align}
    \end{subequations}
    
\subsubsection*{2. Computation stage} The model $\text{SSM}(\overline{\mathbf{A}}, \overline{\mathbf{B}}, \mathbf{C})(.)$ can then be computed as a linear recurrence (Eq.~\ref{eq:s4rnn1}-\ref{eq:s4rnn2}) or as a global convolution (Eq.~\ref{eq:s4rnn3}-\ref{eq:s4rnn4}). 
    \begin{subequations}
    \label{eq:s4rnn}
        \begin{align}
            h_t &= \overline{\mathbf{A}}h_{t-1} + \overline{\mathbf{B}}x_t \label{eq:s4rnn1}\\
            y_t &= \mathbf{C}h_t\label{eq:s4rnn2} \\
            \mathbf{\overline{K}} &= (\mathbf{C}\mathbf{\overline{B}}, \mathbf{C}\mathbf{\overline{A}}\mathbf{\overline{B}}, \dots, \mathbf{C}\mathbf{\overline{A}}^{M-1}\mathbf{\overline{B}}) \label{eq:s4rnn3}\\
            \mathbf{y} &= \mathbf{x} * \mathbf{\overline{K}}\label{eq:s4rnn4}
        \end{align}
    \end{subequations}

Efficient computation of an S4 model necessitates that a structure be imposed on the transition matrix $\mathbf{A}$ since random initialization leads to very poor performance, and thus the name \textit{structured} state space models. The most common form of structure is diagonal, where $\mathbf{A}$ is initialized using the \textit{HiPPO} theory of continuous time memorization \cite{gu2020hippo}. In equations (\ref{eq:s4disc})-(\ref{eq:s4rnn}), matrices $\mathbf{A} \in \mathbb{R}^{N \times N}$, $\mathbf{B} \in \mathbb{R}^{N \times 1}$ and $\mathbf{C} \in \mathbb{R}^{1 \times N}$ can all be represented by $N$ numbers. Furthermore, given that it is an LTI system, the parameters $(\Delta,\mathbf{A},\mathbf{B},\mathbf{C})$ of S4 are fixed for all time-steps. The dual nature of S4 that enables computation as a structured global convolution (Eq.~\ref{eq:s4rnn4}) with a structured kernel $\mathbf{\overline{K}}$ facilitates efficient training, whereas computation via recurrence (Eq.~\ref{eq:s4rnn2}) enables efficient inference, thus offering a ``best of both worlds".


While S4 demonstrated excellent long-term dependency modeling capabilities and achieved state-of-the-art performance on the Long Range Arena \cite{tay2021long} benchmarks, it has several drawbacks. S4, being an LTI system, is not content-aware. More specifically, S4 can neither select correct information from context nor adapt the hidden state based on information presented in a sequence, as evident from the transitions in Eq.~\ref{eq:s4rnn}. In simpler words: once the transition matrices are learned and hence fixed, during inference for a given input, there is no mechanism in S4 to retrieve relevant information for current time-step $t$ based on previous time-step(s), nor is there a mechanism to reflect a relevant change in the hidden state. This is highlighted by empirical evidence provided by \cite{gu2023mamba} that suggests that LTI SSMs struggle at selective copying and context-aware reasoning. Further, when operating on a $B$ sized batch of $D$-channel sequences of length $L$, an SSM is to be applied \textit{independently} on each channel. This is computationally inefficient as it requires applying a total hidden state that has $DN$ dimensions per input and takes $O(BLDN)$ time and memory. 

To address these issues, \cite{gu2023mamba} proposed incorporating a selection mechanism to structured state space models. yielding the ``selective structured state space sequence models with a scan" algorithm, also known as S6. 
S6 trades off time-invariance and equivalence to convolutions for input-dependent processing. The fundamental premise behind S6 is to make SSM parameters $\Delta, \mathbf{B}$ and $\mathbf{C}$ functions of the input through learned linear projections, thus adding an extra \textit{time} dimension to each of these parameters, as depicted in Eq~\ref{eq:mamba}.
\begin{subequations}
\label{eq:mamba}
    \begin{align}
        \mathbf{B} &= s_B(x) \label{eq:mamba1} \\
        \mathbf{C} &= s_C(x) \label{eq:mamba2} \\
        \Delta &= \text{softplus}(\Delta_{\text{p}} + s_{\Delta}(x)) \label{eq:mamba3}
    \end{align}
\end{subequations}
where, $s_B(.)$ and $s_C(.)$ are $N$ dimensional parameterized linear projections, whereas $s_{\Delta}$ is a linear projection to unit dimension broadcasted to all $D$ channels. $\Delta_p$ symbolizes the actual learnable parameter underpining $\Delta$.
Thus, effectively, for a batch of $B$ input sequences of length $L$ and channels $D$, parameters $\mathbf{B}$ and $\mathbf{C}$ are tensors of shape $(B\times L \times N)$, $\Delta$ is a tensor of shape $(B\times L \times D)$ and the resulting discretized parameters $\overline{\mathbf{A}},\overline{\mathbf{B}}$ are tensors of shape $(B\times L \times D \times N)$. 
Finally, the output can be computed using the same $\text{SSM}(\overline{\mathbf{A}},\overline{\mathbf{B}},\mathbf{C})(.)$ formulation as presented earlier, however, since the equivalence to convolution has been lost due to the tensor shapes and input-dependence, it can be computed only in the recurrence mode (Eq.~\ref{eq:s4rnn1}-\ref{eq:s4rnn2}). 
Intuitively, one could look at the S6 architecture, and its constituent parameters and elements, as follows:
\begin{itemize}
    \item The selection mechanism in S6 is closely related to classical RNNs, with the $\Delta$ parameter playing the role of a generalized RNN gating mechanism. $\Delta$ parameter governs the persistence of hidden state $h$ when presented with input $\mathbf{x}$: a large value of $\Delta$ resets $h$ and focuses on the contents of the current input $x_t$, whereas a smaller $\Delta$ results in persistence of the state. 
    \item It is worth noting that the parameter $\mathbf{A}$ is not selective. The reasoning is that it does not need to be, since the model only interacts with $\mathbf{A}$ through $\overline{\mathbf{A}}$, which in turn is impacted by the selective parameter $\Delta$ during the discretization process.
    \item Parameters $\mathbf{B}$ and $\mathbf{C}$ can be further viewed as gates that affect how input $x_t$ impacts the hidden state $h$ ($\mathbf{B}$) and how the hidden state $h$ impacts the output $y_t$ ($\mathbf{C}$). 
\end{itemize}

In the case of S6, the key bottleneck from an implementation standpoint is the complexity of the recurrence computation ($O(BLDN)$). 
While a parallel scan algorithm can be used to implement the recurrence, together with kernel fusion techniques, efficient materializing of the two discretized input tensors $\overline{\mathbf{A}},\overline{\mathbf{B}}$ and the hidden state $h$ is necessary, since operations on these tensors have a considerable memory bandwidth cost.
\begin{figure}[!t]
    \centering
    \includegraphics[width=0.8\linewidth, trim={5.25em 8em 8.5em 2.5em},clip]{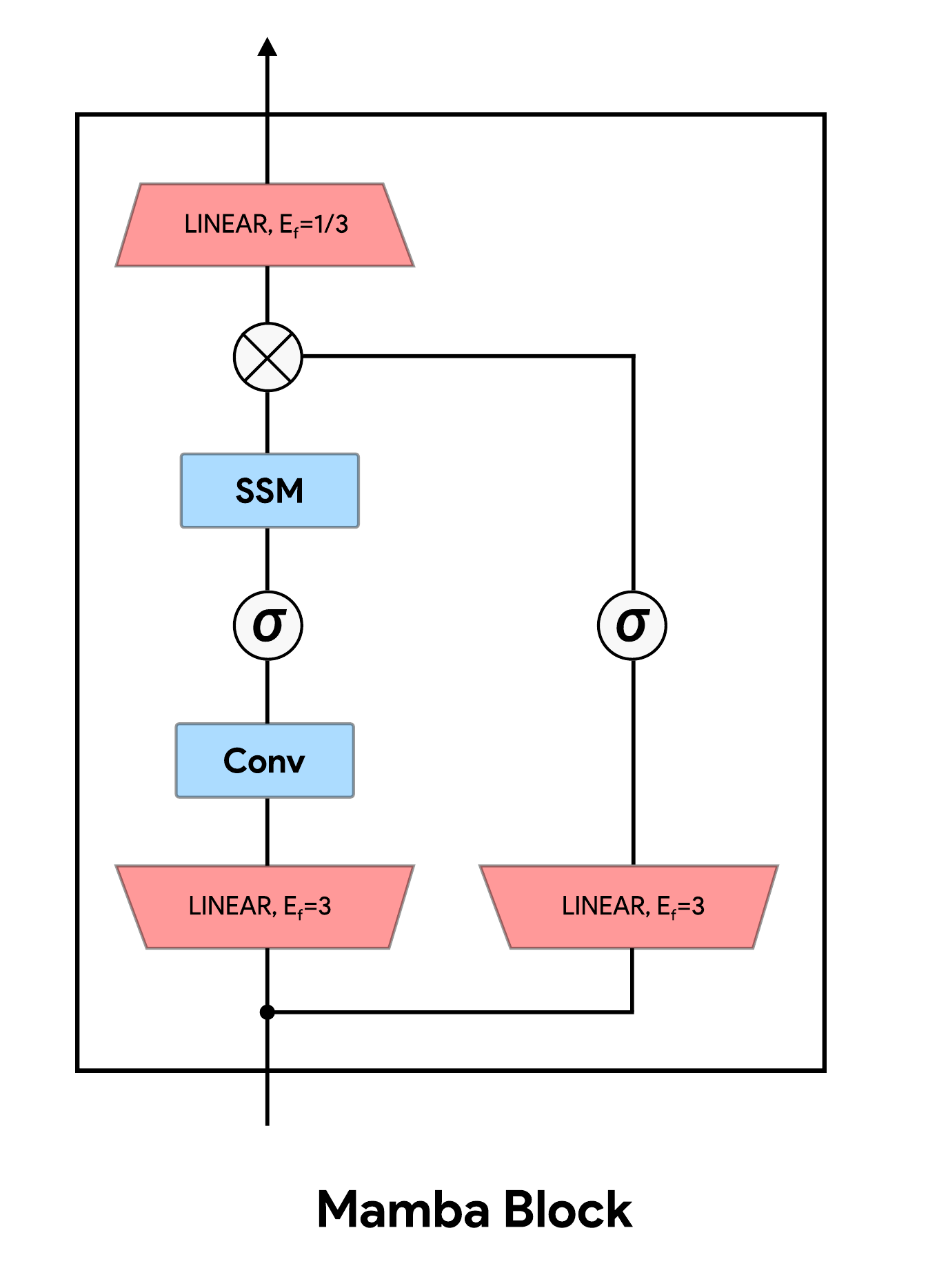}
    \caption{The Mamba block, where SSM represents the Mamba SSM formulation and $\sigma$ signifies the Swish activations. Red blocks are linear projections.}
    \label{fig:mamba}
\end{figure}
To facilitate this, a parallel scan implementation that is aware of the accelerators' memory hierarchy, such that discretization and recurrence occurs directly in faster memory levels of the accelerator (SRAM), with final outputs written directly to slower but larger memory levels (HBM), is developed. Finally, \cite{gu2023mamba} propose a simplified plug-and-play SSM building block, called \textit{Mamba} (Figure~\ref{fig:mamba}), which is a combination of an H3 block \cite{fu2023hungry} and a gated MLP block quite prevalent in modern DNNs. From hereon, we use the term Mamba to refer to both the Mamba blocks and/or S6, depending on the context. Thus, most existing neural architectures can be fitted with an SSM by simply replacing Transformer blocks with Mamba block(s) of appropriate dimensions. Mamba scales linearly with input sequence length, and offers up to 5x better inference throughput compared to Transformers \cite{gu2023mamba}. It can also performs well on up to a million-length sequences, outperforming comparable Transformer models for language modeling, audio and genomics, as demonstrated by \cite{gu2023mamba}.

Consequently, Mamba has gained significant traction across several domains since its release, such as visual recognition \cite{vmambabaa2da9a, visionmamba24}, object recognition and detection \cite{yu2025mambaout}, video understanding \cite{videomamba2025}, point-cloud analysis \cite{pointmamba24, zhang2025point}, DNA sequence modeling \cite{schiff2024caduceus}, biomedical image segmentation \cite{ma2024u}, language modeling \cite{mamba2, mamballama, lenz2025jamba} as well as autoregressive pretraining for foundation models for vision \cite{liu2025map,ren2025autoregressive}. Several derivatives of Mamba have also been proposed, such as Hydra \cite{hwang2024hydra}, which proposes generalized matrix mixers for bidirectional state space modeling, and LocalMamba \cite{localmamba25}, which proposes a windowed selective scan algorithm for better modeling of local dependencies. In the speech and audio processing domain, Mamba has been adopted for speech recognition \cite{yoshiki24, speechmamba, zhang2025mamba, mambastreamingasr25, speechslytherin25, zhang25rethinkingmambaasr}, speech enhancement \cite{luo2024mambagan, semamba24, wang2025mamba, kuhne2025mambattention, chao2025universal, qian2025sav}, speech separation and speaker extraction \cite{li2024spmamba, fan25speakerextraction, avenstrup2025sepmamba, plaquet2025mamba}, depression detection \cite{lin25stemamba,ye25depmamba,liu25mfmamba}, speech super-resolution \cite{lee25waveumamba}, emotion recognition \cite{zhu25mambaser}, music and sound generation \cite{lee2025exploring,chen25musicmamba,colombo25mambafoley}, deepfake detection \cite{chen24k_interspeech, kheir2025bicrossmamba}, audio representation learning and classification \cite{hamza24aum, liang25mamba, mu2024seld, lin2024audio}, and finally, in SSAST-style MSMs for learning self-supervised audio representations \cite{shams2024ssamba, yadav2024audiomambaselectivestate, yuksel2025general}. 

However, Mamba, or other selective state space models for that matter, are not without drawbacks. For instance, Mamba is intrinsically a unidirectional model, which is suboptimal for several tasks and domains. Several works have attempted to address this issue, either by proposing Mamba blocks with additional SSM branch(es) that operate on a flipped version of the input \cite{visionmamba24, yoshiki24}, using multi-dimensional parallel scan operations \cite{vmambabaa2da9a, li2024mamba}, or by matrix mixing \cite{hwang2024hydra}. Using a hybrid Transformer-Mamba approach is also very popular \cite{lenz2025jamba, kuhne2025mambattention, ren2025samba}. It is fair to say that Mamba and state space models have amassed tremendous interest from the community at large. Our understanding of Transformers comes from years of research in the field, and with several challenges yet to be solved and properties yet to be discovered, we believe that state space models such as Mamba are to become a mainstay in signal processing and machine learning.

\subsection{Extended long short-term memory}
\label{ssec:xlstm}

Before the advent of Transformers and state-space models, LSTMs \cite{hochreiter1997long} were the go to neural architecture for sequence modeling and powered several key breakthroughs in neural machine translation \cite{sutskever2014sequence, wu2016google}, automatic speech recognition \cite{graves2006connectionist, amodei2016deep}, image captioning \cite{Karpathy_2015_CVPR, xu2015show}, image generation \cite{gregor2015draw} and audio synthesis \cite{kalchbrenner2018efficient}, to name a few.
Transformers were able to outperform LSTMs and become the de facto sequence modeling architecture in lieu to the following shortcomings of the latter:
\subsubsection*{1. Compression into a scalar cell state} At the heart of the LSTM is a scalar memory cell, which interacts with three gates: input, output and forget gate. All information processed by the LSTM thus has to be compressed and stored in this scalar memory cell, which limits its storage capacity. This can lead to several issues, one of which is poor performance on rare input tokens compared to Transformers, which have no compression of context at all.
\subsubsection*{2. Sigmoid gating and inability to revise storage decisions} In LSTM, the input, output and forget gates have a sigmoid activation for gating. Since sigmoid squashes the input into the $[0.,1.]$ range in an \textit{S}-shaped curve, very large and very small values at the proximities are flattened to values very close to each other. This flattening effect can impact fine-grained decision making regarding storage, and thus LSTM can struggle to revise a stored value when a more similar vector is found.
\subsubsection*{3. Hidden-hidden connections and lack of parallelizability} In LSTMs, hidden-hidden connections between hidden state from one time step to the next, which is also referred to as \textit{memory mixing}, mandates sequential processing. Despite the constant memory and linear computation complexity of the LSTM, in practice this is a bottleneck in training with hardware acceleration. On the other hand, while the SDPA operation has quadratic compute and memory complexity, the inherently parallelizable nature of the SDPA operation facilitates fasting training and better scalability, which was a big win for the Transformer architecture.

The Extended Long Short-Term Memory, a.k.a. xLSTM architecture, \cite{beck2024xlstmextendedlongshortterm} addresses these issues and proposes two core memory cell blocks: sLSTM and mLSTM, both of which encorporate exponential gating on the input and forget gates to enhance storage decision revision capabilities, along with new normalizer and stabilizer states for better training stability.
\begin{figure}[h]
    \centering
    \includegraphics[width=\linewidth, trim={1em 10em 1em 0em},clip]{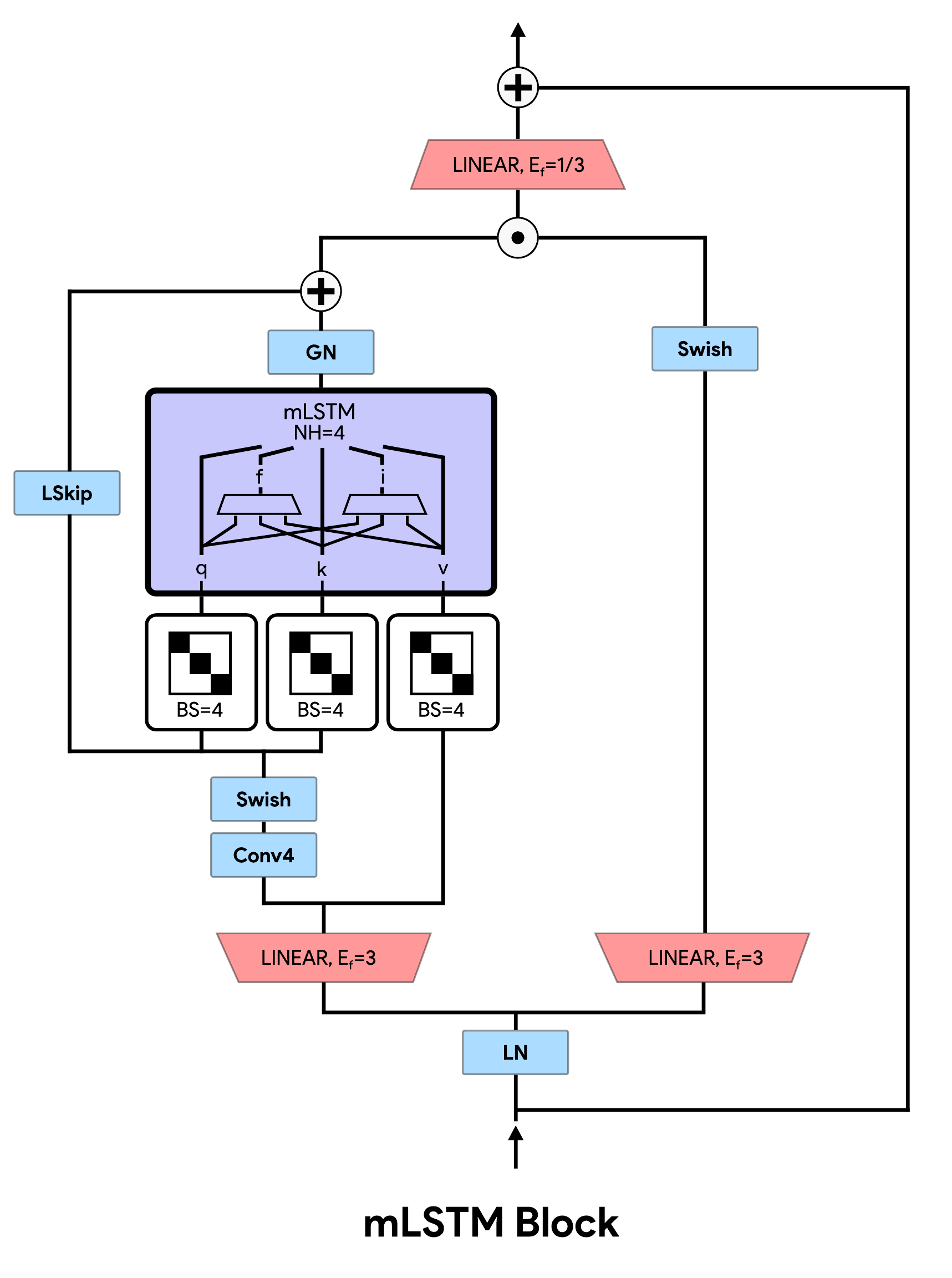}
    \caption{The ViL block fitted with the mLSTM module. GN, LN stand for GroupNorm and LayerNorm, respectively. Red blocks are linear projections.}
    \label{fig:mlstm}
\end{figure}

sLSTM introduces the concept of \textit{heads}, similar to the Transformers, where multiple memory cells can be grouped together in different heads. A normalizer state is introduced, which sums up the product of the input gate times all future forget gates. The scalar sLSTM forward pass is as follows:
\begin{subequations}
\label{eq:slstm}
    \begin{align}
        c_t &= f_t c_{t-1} + i_t z_t \label{eq:slstm1} \\
        n_t &= f_t n_{t-1} + i_t \label{eq:slstm2} \\
        h_t &= o_t \tilde{h}_t, \qquad\tilde{h}_t = c_t / n_t \label{eq:slstm3} \\
        z_t &= \text{tanh}(\textbf{w}_z^t \textbf{x}_t + r_z h_{t-1} + b_z) \label{eq:slstm4} \\
        i_t &= \text{exp}(\textbf{w}_i^t \textbf{x}_t + r_i h_{t-1} + b_i) \label{eq:slstm5} \\
        f_t &= \text{exp}(\textbf{w}_f^t \textbf{x}_t + r_f h_{t-1} + b_f) \label{eq:slstm6} \\
        o_t &= \text{sigmoid}(\textbf{w}_o^t \textbf{x}_t + r_o h_{t-1} + b_o) \label{eq:slstm7}
    \end{align}
\end{subequations}
where $c_t$ is the cell state, $n_t$ is the normalizer state, $h_t$ is the hidden states, $z_t$ is the cell input and $i_t, f_t, o_t$ represent the input, forget and the output gate, respectively. Weight vectors between the input $\mathbf{x}_t$ and the corresponding unit/gate are represented as $\mathbf{w}_*$, whereas recurrent weights between hidden state $h_{t-1}$ and the corresponding unit/gate are represented as $r_*$. $b_*$ represent bias terms. Since exponential activations ($\text{exp}(.)$) can lead to large values that can cause overflows, input and forget gates are stabilized as follows:
\begin{subequations}
\label{eq:slstmstab}
    \begin{align}
        m_t &= \text{max}(\log(f_t) + m_{t-1}, \log(i_t)) \label{eq:slstmstab1} \\
        i'_t &= \text{exp}(\log(i_t) - m_t) \label{eq:slstmstab2} \\
        f'_t &= \text{exp}(\log(f_t) + m_{t-1} - m_t) \label{eq:slstmstab3} 
    \end{align}
\end{subequations}
where $m_t$ denotes the stabilizer state. In lieu to a new memory mixing strategy that enables inter-head memory mixing, i.e., hidden-hidden connections between time steps within the constituent memory cells in a head, but no intra-head mixing, sLSTM allow sequential-only processing, and are closer to the original LSTM.

mLSTM enhances the storage capacity of LSTMs by replacing the scalar memory cell $c \in \mathbb{R}$ in the LSTM with a matrix cell $\mathbf{C} \in \mathbb{R}^{d \times d}$. Eq.~\ref{eq:mlstm} depicts the forward pass for an mLSTM block for an input $\mathbf{x}_t$. Note that as opposed to sLSTM, states are vectors.
\begin{subequations}
\label{eq:mlstm}
    \begin{align}
    \mathbf{C}_t &= f_t \mathbf{C}_{t-1} + i_t \mathbf{v}_t \mathbf{k}_t^T \label{eq:mlstm1} \\
    \mathbf{n}_t &= f_t \mathbf{n}_{t-1} + i_t \mathbf{k}_t \label{eq:mlstm2} \\
    \mathbf{h}_t &= \mathbf{o}_t \odot \left(\frac{\mathbf{C}_t \mathbf{q}_t}{\max(|\mathbf{n}_t^T \mathbf{q}_t|, 1)}  \right) \label{eq:mlstm3} \\
    \mathbf{q}_t &= \mathbf{W}_q \mathbf{x}_t + \mathbf{b}_q \label{eq:mlstm4} \\
    \mathbf{k}_t &= \frac{1}{\sqrt{d}}\mathbf{W}_k \mathbf{x}_t + \mathbf{b}_k \label{eq:mlstm5} \\
    \mathbf{v}_t &= \mathbf{W}_v \mathbf{x}_t + \mathbf{b}_v \label{eq:mlstm6} \\
    i_t &= \text{exp}(w_i^T x_t + b_i) \label{eq:mlstm7} \\
    f_t &= \text{exp}(w_f^T x_t + b_f) \label{eq:mlstm8} \\
    \mathbf{o}_t &= \text{sigmoid}(\mathbf{W}_o \mathbf{x}_t + \mathbf{b}_o) \label{eq:mlstm9} 
    \end{align}
\end{subequations}
Aligning with the terminology introduced by Transformers, mLSTMs computes query ($\mathbf{q}$), key ($\mathbf{k}$), and value ($\mathbf{v}$) vectors from the input vector $\mathbf{x}$, also introducing corresponding weight matrices ($\mathbf{W}_q,\mathbf{W}_k, \mathbf{W}_v$, respectively) and bias terms. The core memory cell $\mathbf{C}$ is formed as the outer product between the key and the value vectors, abiding with the covariance update rule given that keys and values are expected to have zero mean since projection is proceeded by a layer-norm.
In contrast to the sLSTM, multiple memory cells and heads are equivalent in an mLSTM since there is no memory mixing, and thus the mLSTM cell can be parallelized. mLSTM uses the same stabilization as sLSTM (Eq.~\ref{eq:slstmstab}). \cite{beck2024xlstmextendedlongshortterm} propose two different residual blocks for the two types of cells. The sLSTM follows a \textit{post-up projection} like Transformers, where sequence modeling is done in the original space followed by projection to a higher dimensional space and applying the non-linearity, whereas the mLSTM block adopts a \textit{pre-up projection} block similar to Mamba where projection to a higher dimensional space is done before sequence modeling, facilitating a larger memory capacity. The xLSTM architecture, thus, is a stack of these sLSTM and/or mLSTM blocks. 
xLSTMs have demonstrated excellent performance for language modeling, performing on par or better than Transformers \cite{beck2024xlstmextendedlongshortterm, beck2025xlstm}. While being more recent compared to state space models, xLSTMs have been leveraged in a variety of domains, such as ViL, a patch based image recognition approach similar to Vision Transformers \cite{alkin2024visionlstmxlstmgenericvision} which leverages only the mLSTM blocks, as well as generative and in-context learning of DNA, protein and chemical sequences \cite{schmidinger2025bioxlstm}. 
xLSTMs are also seeing active adoption in remote sensing \cite{zhu2024seg}, time-series modeling \cite{kraus2024xlstm, kong2025unlocking, auer2025tirex} and medical image segmentation \cite{heidari2025study,zhu2025xlstm,chen2024xlstm,dutta2024vision}, majority of latter building on top of ViL blocks. In the audio and signal processing domain, xLSTMs have being used for speech enhancement \cite{kuhne2025xlstm}, and finally, in SSAST-style MSMs for learning self-supervised audio representation \cite{yadav2025axlstms}, which also uses ViL based blocks that comprise primarily of mLSTM cells. Thus, it is evident that between the two proposed blocks, the mLSTM block has definitely garnered much more attention, and is utilized in upcoming sections of this paper. Figure~\ref{fig:mlstm} depicts the mLSTM based ViL block.

The xLSTM architecture has it's own set of flaws. Until very recently, xLSTMs were much slower in practice compared to Transformers, despite their linear complexity with respect to input sequence lengths. This was recently addressed in \cite{beck2025tiled}, where more performant GPU kernels were proposed. Similar to Mamba, xLSTMs are also intrinsically a unidirectional model, and requires alternating blocks that operate on flipped input for bidirectional modeling \cite{alkin2024visionlstmxlstmgenericvision}. Further, from practical real world usage, xLSTMs have been known to require more memory compared to Transformers for short length sequences. Nevertheless, despite being a more recent development than Mamba, xLSTMs are seeing active adoption in the field, and are a promising research direction in signal processing and machine learning.

In the upcoming sections we aim to provide an empirical comparison of Transformers, Mamba and xLSTMs for masked spectrogram modeling across a wide variety of downstream audio recognition tasks within a consistent, reproducible unified MSM framework.

\section{Approach}
\label{sec:approach}

To evaluate Transformers, Mamba and xLSTM under a unified framework, we adopt an SSAST-style MSM. We chose SSAST over MAE for a critical reason: encoding as done in MAE is closely coupled and only possible with the Transformer architecture because it treats input as a set of tokens and assumes no sequential relationship between consecutive patches. Picking a framework that is too closely tied with a particular sequence modelling approach prohibits evaluation of alternative, upcoming approaches. MAE is inherently incompatible with Mamba and xLSTM, which are fundamentally recurrent neural networks that are explicitly modelling sequential dynamics in the input. In contrast, the SSAST design is not coupled with the Transformer architecture, and can be used to evaluate all kinds of sequence modelling approaches. Thus, the only way to conduct an unbiased comparison of Transformers with Mamba and xLSTM without additional confounding factors was to do so within the SSAST framework. SSAST was the first patch-based self-supervised learning framework for audio spectrogram Transformers, and it is a well established baseline that will facilitate comparisons across a multitude of application domains. In the upcoming section we first discuss the datasets used in the study, followed by a description of the unified MSM framework used for comparison. Finally, we discuss the post-pretraining downstream evaluation setup.

\subsection{Datasets}
\label{ssec:approach_datasets}

\subsubsection*{Pretraining} We use the AudioSet dataset \cite{gemmeke2017audio} for self-supervised pretraining. AudioSet is one of the largest audio corpus available, with roughly 2-million 10-second long weakly labeled YouTube video clips that span an exhaustive hierarchy of over 527 labels, totalling over 5000 hours of audio data. The choice for AudioSet was straightforward: it is a prominent dataset widely used for training audio representations \cite{kong2020panns, saeed2021contrastive, niizumi2021byol}, and almost all popular MSMs in literature have used AudioSet \cite{niizumi2022masked, huang2022masked, yadav2024masked, gong2022ssast}. 

\subsubsection*{Downstream Evaluation} The objective of self-supervised pretraining for audio is to create a general purpose audio representation that performs well across a wide variety of audio recognition tasks. There are innumerous well-established audio recognition datasets in existence across a wide variety of domains, such as keyword spotting \cite{warden2018speech}, environmental audio classification \cite{esc50}, audio tagging \cite{fonseca2021fsd50k, gemmeke2017audio} and emotion recognition \cite{cao2014crema, busso2008iemocap}, to name a few. Furthermore, several benchmark suites have emerged to evaluate the performance of audio representations on a wide swath of datasets, such as SUPERB \cite{yang21c_interspeech} for evaluating speech representations. These benchmark suites consolidate datasets that cover a wide variety of downstream tasks and provide standardized and reproducible evaluation protocols. HEAR \cite{turian_hear_2022} is one such suite. HEAR covers over 19 downstream audio recognition tasks, including a mix of established prominent audio recognition benchmarks as well as new ones, and has been used by several MSMs for evaluation \cite{niizumi2022masked, yadav2024masked, dinkel24b_interspeech}. However, several of these tasks are either redundant or highly correlated in performance \cite{turian_hear_2022}. Thus, similar to \cite{yadav2024masked}, we utilize a 10-task subset of the HEAR benchmark, for evaluating our MSM, which offers the best tradeoff between sufficient variety in downstream tasks while limiting the number of evaluations conducted. More specifically, for evaluating music and pitch perception characteristics of the evaluated MSMs, we use Beijing Opera \cite{beijingopera, turian_hear_2022}, Mridangam Stroke and Tonic \cite{mridangamds} and the NSynth Pitch-5h \cite{turian_hear_2022, nsynth2017} tasks. Further, we use Crema-D \cite{cao2014crema}, Speech Commands 5h \cite{turian_hear_2022, warden2018speech}, LibriCount \cite{stoter2018classification} and VoxLingua107 \cite{kim2018vocal} to evaluate MSM performance on speech based tasks. Finally, we use ESC-50 \cite{esc50} and FSD50K \cite{fonseca2021fsd50k} to evaluate general audio classification performance. More information about the downstream tasks can be found in Table~\ref{tab:tasks}, and in their respective references.
\begin{table}[h]
  \caption{Description of the evaluated downstream tasks. \#Classes and \#Hours represent the number of classes and the total duration in hours, respectively.}
  \setlength\tabcolsep{2pt}
  \label{tab:tasks}
  \centering
  \scriptsize
  \begin{tabular}{lcccc}
    \toprule
    ID & Name     & Description   & \#Classes & \#Hours  \\
    \midrule
    BO & Beijing Opera  & percussion instrument classification & 4 & 0.3 \\
    CD & Crema-D    & emotion recognition & 6 & 10 \\
    E50 & ESC-50  & environmental sound classification & 50 & 2.77 \\
    LC & LibriCount & counting speakers (classification) & 10 & 8  \\
    Mri-S & Mridangam Stroke & classifying Mridangam \textit{strokes} & 10 & 1.57  \\
    Mri-T & Mridangam Tonic  & classifying Mridangam \textit{tonics} & 6 & 1.57  \\
    NS-5h & NSynth Pitch 5h  & pitch classification & 88 & 5.5  \\
    SC-5h & SpeechCommands 5h  & keyword spotting & 12 & 6.5  \\
    F50K & FSD50K & multilabel audio tagging & 200 & 100  \\
    VL & VoxLingua107 Top10 & spoken language identification & 10 & 5  \\
    \bottomrule
  \end{tabular}
\end{table}

\subsection{Model architecture}
\label{ssec:approach_arch}

To compare Transformers, Mamba and xLSTM, we utilize an SSAST-MSM framework as specified in Section~\ref{sec:msm}, and then essentially replace the Transformer blocks with Mamba blocks as specified Section~\ref{ssec:ssms} and Figure~\ref{fig:mamba} and xLSTM blocks as specified in Section~\ref{ssec:xlstm} and Figure~\ref{fig:mlstm}. The resulting models with Transformer, Mamba and xLSTM based sequential modeling are hereafter referred to as SSAST, Self-Supervised Audio Mamba (SSAM) \cite{yadav2024audiomambaselectivestate} and Audio-xLSTMs (AxLSTM) \cite{yadav2025axlstms}, respectively. In this section, we elucidate the specifics of the adopted SSAST-style MSMs across the Transformer, Mamba and xLSTM sequence modeling paradigms.

\subsubsection*{Creating and masking patches} Following \ref{ssec:patching} and \ref{ssec:maskingstrats}, all the models follow identical patch creation and masking. 
Log-scaled mel spectrogram features extracted from audio clips sampled at 16000 Hz with a window size of 25 ms, a hop size of 10 ms and F = 80 mel-spaced frequency bins are used as inputs, and in our default configuration we use a 2-second random audio crop for self-supervised pretraining. Unless specified otherwise, we compute (4x16) shaped rectangular patches, i.e. our patch parameters are $t=4,f=16$. This corresponds to spectrogram input to the MSM $\mathbf{X} \in \mathbb{R}^{200 \times 80}$, which after creating $t=4,f=16$ sized patches yields $\mathbf{X}_p \in \mathbb{R}^{250 \times 64}$, which are projected to $d_{\text{enc}}$ (covered in the next paragraph).
Like majority of the MSMs, we use unstructured random masking, along with a masking ratio of $m_r=0.5$ for all models, which is in line with \cite{gong2022ssast}. We add a representative $\text{cls}$ token for consistent comparison with previous works as well as ease of finetuning for future works that use our released models. A fixed sinusoidal embedding is used for encoding positional information.

\subsubsection*{Encoding and reconstruction} The encoder in our default configuration for all the models match up with the ViT-Tiny \cite{dosovitskiy2021an} specifications, i.e. a stack of $l=12$ blocks, $d_{\text{enc}}=192$ and number of attention heads, $h=3$. We also evaluate Small ($d_{\text{enc}}$=$384$, $l$=$12$, $h$=$6$) and Base ($d_{\text{enc}}$=$768$, $l$=$12$, $h$=$12$) encoder configurations across the board. As evident from Figure~\ref{fig:mamba}, Mamba blocks apply the SSM on expanded input, before projecting it back to $d_\text{enc}$ dimensions. Similarly, the mLSTM based ViL block also expands the input (Figure~\ref{fig:mlstm}). For Mamba and mLSTM blocks, the default expansion factor is $E_f=2$, which results in roughly every 2 Mamba/mLSTM blocks having similar number of parameters as a single Transformer block. However, using these settings as is would mean that all Mamba/xLSTM based models were 2 times deeper than the corresponding Transformer based model. To control for neural network depth, which has been shown to directly impact the modeling capabilities of neural networks \cite{szegedy2015going}, we use an expansion factor of $E_f=3$ for both mLSTM-based ViL blocks and the Mamba block. Additionally, the Mamba block also uses larger internal dimensions ($d_{state}=24,d_{conv}=4$). These choices, while still resulting in Mamba/xLSTM based MSMs to have fewer parameters than their Transformer counterparts, allow us to close the gap while controlling for model depth. Furthermore, neither Mamba nor xLSTM based MSMs are bi-directional, as bi-directional processing did not improve performance as per ablations. After encoding, a single hidden layer MLP is used for reconstruction, where the hidden layer has dimensions $d_\text{dec} = d_{\text{enc}}$ and the final layer has dimensions $(t.f)$, with a GELU non-linear activation after the hidden layer.

\subsubsection*{Optimization} As highlighted in Section~\ref{ssec:decoder}, the objective function for pretraining all the MSM is mean squared error (MSE) between the \textit{masked} spectrogram patches and the reconstructions. This is the key difference between all the evaluated MSMs and the original SSAST \cite{gong2022ssast} which utilized the InfoNCE loss function in addition to MSE. All the evaluated models are pretrained for a 100 epochs with a batch size of 1024
and a weight decay of 0.05 with the AdamW optimizer \cite{loshchilov2018decoupled}. A linear warmup for 10 epochs followed by a cosine learning rate decay schedule is used. In accordance with several MSMs in literature \cite{niizumi2022masked, gong2022ssast, huang2022masked, yadav2024masked}, no data augmentations were used. 

\subsection{Downstream evaluation and performance metrics}

As previously stated, the HEAR has a standardized and reproducible evaluation protocol, implemented and released in the public \textit{hear-eval-kit}. We evaluate MSMs on downstream tasks in accordance with the HEAR protocol: we extract fixed-sized feature vectors independent of the input audio duration by taking the mean over time across 2-second audio chunks. Then, a single hidden layer MLP classifier with 1024 neurons is trained for each task, repeated 10 times for each evaluated model. We then report 95\% confidence intervals based on these runs. This setting represents a restricted hyperparameter evaluation grid compared to the one used to obtain results in \cite{turian_hear_2022}, which might cause some discrepancies in numbers reported for the same task between \cite{turian_hear_2022} and ours. However, this was necessary given the large number of experiments conducted, and is in line with existing work \cite{yadav2024audiomambaselectivestate,yadav2024masked, yadav2025axlstms}). While these results differ from a full fine-tuning use case, this evaluation protocol results in a thorough qualitative assessment of the evaluated self-supervised audio representations, where fine-tuning performance would be more bound by the number of parameters of the model, and not to mention, incredibly inefficient and wasteful of compute resources.

\subsubsection*{Aggregated Performance Metric} While all the evaluated downstream tasks are utterance level classification tasks, they are all vary significantly in difficulty and \textit{breadth} in the underlying objective.  Furthermore, different self-supervised representations will naturally do better on some tasks versus the other. Additionally, keeping track of performance across all the tasks is arduous, especially for ablations. Thus, using an aggregated performance metric to quantify the performance of an evaluated self-supervised model across all the tasks, that also takes task complexity into account, would be incredibly useful. To this end, we use the aggregated normalized score as proposed by \cite{yadav2024masked} to compare evaluated approaches across the proposed list of downstream tasks. For a model $m$, overall score $s(m) \in [0,100]$ is given as:
\begin{equation}
    \label{eq:sm}
    s(m) = \frac{1}{|T|}\sum_{t \in T} \frac{x_t(m) - \text{min}_t}{\text{max}_t - \text{min}_t} * 100
\end{equation}
where $x_t(m)$ denotes performance of the model $m$ on task $t$, and $\text{min}_t$ and $\text{max}_t$ represent the worst and the best performance across all models on the task, thus taking into account the relative performance amongst all evaluated representations. This metric is similar to the one proposed by the SUPERB leaderboard \cite{yang21c_interspeech}, with a difference in the scale and range of the metric.

\section{Empirical analysis}
\label{sec:emp}

\subsection{Establishing contextual performance w.r.t. literature}
While our primary objective is to compare Transformers, Mamba and xLSTM based MSMs against each other, it is necessary to contextualize their performance compared to representations in literature. To this end, alongside pretraining SSAST-style Transformer, Mamba and xLSTM MSMs, we also conducted downstream experiments on several prominent audio representations, as shown in Table~\ref{tab:overall}. As a supervised baseline, we used the Patchout faST Spectrogram Transformer (PaSST) \cite{koutini22_interspeech}. Trained on the AudioSet data in a supervised manner, PaSST establishes a very strong audio recognition baseline, improving upon plain Audio Spectrogram Transformers (ASTs) \cite{gong21b_interspeech} by a considerable margin. We also include prominent self-supervised speech representations: wav2vec 2.0 \cite{baevski2020wav2vec}, HuBERT \cite{hsu2021hubert} and WavLM \cite{chen2022wavlm} in our analysis, since these approaches have seen widespread adoption and are a useful landmark to navigate self-supervised audio representation performance. Finally, we include several popular existing MSMs, including the original SSAST \cite{gong2022ssast} and other MAE-style MSMs \cite{huang2022masked, niizumi2022masked, chenbeats23, yadav2024masked}, to better contextualize performance of evaluated MSMs.

\begin{table*}[t]
    \caption{Comparing evaluated MSMs with popular self-supervised audio representations. LS, AS, VP, LL stand for LibriSpeech, AudioSet, VoxPopuli and LibriLight datasets, respectively. Original SSAST \cite{gong2022ssast} was trained on AS+LS, whereas we pretrained the directly comparable underlined SSAST baselines (SSAST-Tiny,Small,Base). *includes decoder parameters}
    \setlength\tabcolsep{2.25pt}
    \small
    \centering
    \begin{tabular}{lcc|cccc|cccc|cc|c}
    \toprule
    \multicolumn{3}{l}{} &\multicolumn{4}{c}{Music \& Pitch}  &\multicolumn{4}{c}{Speech-based tasks} &\multicolumn{2}{c}{Audio} & \\
    \cmidrule(lr){4-7} \cmidrule(lr){8-11} \cmidrule(lr){12-13}
    \multicolumn{1}{l}{Model} & \multicolumn{1}{l}{Data} & \#M Params & BO & Mri-S & Mri-T & NS-5h & CD & LC & SC-5h & VL & E50 & F50K & $s(m)$ \\
    \midrule
    \multicolumn{3}{l}{\textbf{Supervised Baselines}} & \multicolumn{5}{l}{}\\
    \rowcolor{Gray}
    HEAR-Naive \cite{turian_hear_2022} & - & - & 52.6\ci{2.4} & 38.0\ci{1.3} & 36.4\ci{1.9} & 18.6\ci{4.4} & 30.9\ci{0.8} & 33.5\ci{1.1} & 8.5\ci{0.4} & 11.2\ci{0.5} & 5.8\ci{0.2} & 7.1\ci{0.2} & 5.1\ci{0.7}\\
    PaSST-Base \cite{koutini22_interspeech} & AS & $86$ & 94.9\ci{0.5} & 96.5\ci{0.1} & 87.6\ci{0.6} & 23.3\ci{0.9} & 61.0\ci{0.3} & 60.1\ci{0.2} & 66.6\ci{1.4} & 25.5\ci{0.8} & \textbf{94.8\ci{0.3}} & \textbf{64.2\ci{0.1}} & 73.7\ci{0.4}\\

    \midrule
    \multicolumn{3}{l}{\textbf{SSL}} & \multicolumn{5}{l}{}\\
    \rowcolor{Gray}
    W2V2-base \cite{baevski2020wav2vec} & LS & $94.4$ & 74.0\ci{1.0} & 77.3\ci{0.2} & 55.1\ci{0.3} & 7.4\ci{0.8} & 46.4\ci{0.3} & 51.2\ci{0.2} & 90.8\ci{0.3} & 35.5\ci{0.8} & 31.1\ci{0.4} & 18.1\ci{0.1} & 43.2\ci{0.2}\\
    W2V2-large \cite{baevski2020wav2vec} & VP & $315.4$ & 93.1\ci{0.7} & 93.9\ci{0.1} & 77.4\ci{0.2} & 42.0\ci{1.0} & 66.9\ci{0.4} & 62.4\ci{0.3} & 87.6\ci{0.5} & 53.6\ci{1.0} & 60.1\ci{0.5} & 34.2\ci{0.1} & 74.2\ci{0.4}\\
    \rowcolor{Gray}
    HuBERT-base \cite{hsu2021hubert} & LS & $94.4$ & 92.1\ci{0.6} & 94.4\ci{0.1} & 84.9\ci{0.3} & 19.4\ci{0.7} & 70.8\ci{0.2} & 56.5\ci{0.3} & 93.2\ci{0.1} & 61.8\ci{0.6} & 57.8\ci{0.6} & 32.3\ci{0.1} & 72.7\ci{0.2}\\
    HuBERT-large \cite{hsu2021hubert} & LL & $315.4$ & 94.1\ci{0.7} & 95.3\ci{0.1} & 83.5\ci{0.3} & 19.3\ci{0.8} & 70.7\ci{0.1} & 59.9\ci{0.2} & 83.2\ci{0.7} & \textbf{66.1\ci{0.9}} & 60.3\ci{0.4} & 31.5\ci{0.1} & 73.6\ci{0.3}\\
    \rowcolor{Gray}
    WavLM-base \cite{chen2022wavlm} & LS & $94.4$ & 89.4\ci{0.7} & 95.1\ci{0.1} & 83.4\ci{0.2} & 37.3\ci{0.8} & 56.3\ci{0.2} & 63.2\ci{0.3} & 57.2\ci{0.8} & 22.6\ci{0.6} & 46.6\ci{0.4} & 29.9\ci{0.1} & 60.7\ci{0.2}\\
    WavLM-large \cite{chen2022wavlm} & Mix & $315.4$ & \textbf{96.4\ci{0.5}} & 96.8\ci{0.1} & 89.5\ci{0.1} & 53.7\ci{0.5} & 57.2\ci{0.2} & 61.1\ci{0.3} & 46.2\ci{0.8} & 23.7\ci{0.9} & 47.9\ci{0.4} & 29.0\ci{0.1} & 64.2\ci{0.2}\\
    \midrule
    \multicolumn{3}{l}{\textbf{MAE-Style MSMs}} & \multicolumn{5}{l}{}\\
    \rowcolor{Gray}
    AudioMAE \cite{huang2022masked} & AS & $86.0$ & 93.7\ci{0.6} & 89.2\ci{0.2} & 86.6\ci{0.2} & 64.5\ci{0.8} & 68.2\ci{0.2} & 42.2\ci{0.2} & 28.6\ci{1.5} & 29.7\ci{1.0} & 60.6\ci{0.4} & 37.9\ci{0.1} & 63.1\ci{0.3}\\
    MSM-MAE-208 \cite{niizumi2022masked} & AS & ${92.7}^{*}$ & 95.7\ci{0.7} & 97.3\ci{0.1} & 97.9\ci{0.1} & 69.1\ci{0.5} & 68.7\ci{0.2} & 63.8\ci{0.5} & 85.7\ci{0.3} & 40.3\ci{0.6} & 78.4\ci{0.6} & 49.5\ci{0.1} & 85.3\ci{0.2}\\
    \rowcolor{Gray}
    MWMAE-Tiny \cite{yadav2024masked} & AS & ${12.6}^{*}$ & 93.3\ci{1.0} & 97.1\ci{0.1} & 97.6\ci{0.1} & 68.1\ci{0.4} & 64.4\ci{0.2} & 65.5\ci{0.3} & 77.0\ci{0.6} & 28.6\ci{1.1} & 71.9\ci{0.5} & 43.4\ci{0.1} & 79.3\ci{0.3}\\
    
    MWMAE-Base \cite{yadav2024masked} & AS & ${92.5}^{*}$ & 96.0\ci{0.5} & 97.4\ci{0.1} & 97.9\ci{0.1} & 69.3\ci{0.6} & 73.1\ci{0.3} & 68.8\ci{0.2} & 90.9\ci{0.2} & 44.2\ci{0.9} & 81.2\ci{0.4} & 51.2\ci{0.2} & 89.4\ci{0.2}\\
    \rowcolor{Gray}
    MWMAE-Large \cite{yadav2024masked} & AS & ${308.9}^{*}$ & 95.9\ci{0.5} & 97.4\ci{0.0} & \textbf{98.2\ci{0.1}} & 71.2\ci{0.7} & \textbf{76.1\ci{0.2}} & \textbf{69.7\ci{0.3}} & \textbf{93.0\ci{0.1}} & 51.9\ci{0.7} & 83.6\ci{0.3} & 53.5\ci{0.1} & \textbf{92.6\ci{0.2}}\\
    BEATs-iter3 \cite{chenbeats23} & AS & $90.0$ & 94.0\ci{0.8} & 94.7\ci{0.1} & 95.8\ci{0.1} & 69.4\ci{0.8} & 67.3\ci{0.2} & 68.0\ci{0.2} & 85.2\ci{0.3} & 38.5\ci{1.0} & 83.7\ci{0.3} & 53.6\ci{0.2} & 85.9\ci{0.3}\\
    \midrule
    \multicolumn{3}{l}{\textbf{SSAST-Style MSMs}} & \multicolumn{5}{l}{}\\
    \rowcolor{Gray}
    SSAST \cite{gong2022ssast} & Mix & $89.0$ & 93.4\ci{0.9} & 96.7\ci{0.1} & 96.3\ci{0.1} & 66.8\ci{0.7} & 56.5\ci{0.2} & 60.7\ci{0.3} & 53.5\ci{1.3} & 28.5\ci{0.9} & 68.4\ci{0.4} & 38.2\ci{0.1} & 71.9\ci{0.2}\\
    \underline{SSAST-Tiny} & AS & $5.4$ & 90.4\ci{0.7} & 95.7\ci{0.1} & 94.3\ci{0.1} & 61.2\ci{0.5} & 46.9\ci{0.2} & 42.7\ci{0.2} & 50.6\ci{1.6} & 13.8\ci{1.0} & 42.4\ci{0.6} & 24.6\ci{0.1} & 55.3\ci{0.2}\\
    \rowcolor{Gray}
    SSAM-Tiny \cite{yadav2024audiomambaselectivestate} & AS & $4.8$ & 93.7\ci{0.8} & 97.1\ci{0.1} & 94.9\ci{0.1} & 62.0\ci{0.7} & 61.8\ci{0.3} & 59.2\ci{0.4} & 74.8\ci{0.4} & 27.8\ci{1.0} & 70.6\ci{0.2} & 41.3\ci{0.2} & 75.0\ci{0.2}\\
    AxLSTM-Tiny \cite{yadav2025axlstms} & AS & $4.3$ & 93.9\ci{0.7} & 96.8\ci{0.1} & 94.9\ci{0.1} & 62.5\ci{0.6} & 57.5\ci{0.2} & 55.1\ci{0.5} & 69.0\ci{1.6} & 24.1\ci{0.6} & 61.3\ci{0.6} & 37.5\ci{0.2} & 70.1\ci{0.2}\\
    \rowcolor{Gray}
    \underline{SSAST-Small} & AS & $21.5$ & 93.2\ci{0.5} & 96.2\ci{0.1} & 95.0\ci{0.1} & 63.8\ci{0.4} & 51.6\ci{0.2} & 50.0\ci{0.3} & 58.3\ci{1.2} & 15.6\ci{0.7} & 50.1\ci{0.6} & 31.6\ci{0.1} & 62.5\ci{0.2}\\
    SSAM-Small \cite{yadav2024audiomambaselectivestate} & AS & $17.9$ & 94.0\ci{0.7} & 97.5\ci{0.1} & 96.7\ci{0.1} & 66.3\ci{0.8} & 67.5\ci{0.2} & 60.5\ci{0.3} & 83.7\ci{0.3} & 39.6\ci{0.7} & 78.7\ci{0.6} & 48.5\ci{0.1} & 82.8\ci{0.3}\\
    \rowcolor{Gray}
    AxLSTM-Small \cite{yadav2025axlstms} & AS & $16.7$ & 92.9\ci{1.0} & 97.4\ci{0.1} & 96.6\ci{0.1} & 66.6\ci{0.4} & 65.0\ci{0.2} & 60.3\ci{0.3} & 80.5\ci{0.4} & 36.5\ci{0.7} & 75.5\ci{0.4} & 46.5\ci{0.1} & 80.4\ci{0.3}\\
    \underline{SSAST-Base} & AS & $85.7$ & 93.1\ci{0.7} & 96.6\ci{0.1} & 96.2\ci{0.2} & 64.6\ci{0.8} & 56.0\ci{0.4} & 52.9\ci{0.3} & 66.1\ci{1.0} & 19.2\ci{0.9} & 59.6\ci{0.7} & 37.5\ci{0.1} & 68.2\ci{0.3}\\
    \rowcolor{Gray}
    SSAM-Base \cite{yadav2024audiomambaselectivestate} & AS & $69.3$ & 93.2\ci{1.1} & \textbf{97.7\ci{0.1}} & 96.9\ci{0.1} & 70.5\ci{0.5} & 70.3\ci{0.2} & 63.5\ci{0.2} & 87.9\ci{0.3} & 50.4\ci{0.7} & 81.0\ci{0.3} & \underline{52.2\ci{0.1}} & 87.9\ci{0.3}\\
    AxLSTM-Base \cite{yadav2025axlstms} & AS & $65.6$ & 93.6\ci{0.9} & 97.5\ci{0.1} & 97.5\ci{0.1} & 71.4\ci{0.8} & 68.7\ci{0.2} & 63.2\ci{0.3} & 85.1\ci{0.2} & 43.5\ci{0.6} & 79.2\ci{0.6} & \underline{51.0\ci{0.1}} & 85.8\ci{0.2}\\
    \bottomrule
    \end{tabular}
    \label{tab:overall}
\end{table*}
In Table~\ref{tab:overall}, the directly comparable Transformer (SSAST, underlined), Mamba (SSAM) and xLSTM (AxLSTM) approaches can be found in the ``ours" section, and these models have identical feature embedding sizes. Given that it is infeasible to retrain all existing pretrained representations to have the same embedding sizes, we had to use existing representations as is. While it might seem suboptimal, this is a well-established evaluation protocol for comparing self-supervised audio representations \cite{turian_hear_2022, yang21c_interspeech}.
It is worth noting that \textit{SSAST \cite{gong2022ssast}} represents the officially released model trained on AudioSet+LibriSpeech, whereas the underlined SSAST models represent the architecture highlighted in Section~\ref{ssec:approach_arch} which was pretrained by us. As evident from the table, SSAST-based MSMs fitted with Mamba (SSAM) and xLSTM (AxLSTM) blocks outperform comparable Transformer based SSAST baselines by a considerable margin in overall score $s(m)$.  For the evaluated SSAST-style MSMs, the general performance trend indicates that Mamba-based SSAMs $>$ xLSTM-based AxLSTMs $>>$ SSASTs. SSAMs consistently outperform SSASTs by roughly $30\%$ in relative performance, whereas they are only $3-4\%$ better than comparable AxLSTMs counterparts. Both SSAM-Base and AxLSTM-Base outperform MAE-style approaches that are based on the the standard Transformer block like AudioMAE, BEATs-iter3 and MSM-MAE-208, while having fewer parameters. Only the MWMAE \cite{yadav2024masked} models, which possess a modified, non-standard multi-head attention module explicitly designed to capture local-global attention are able to outperform SSAM and AxLSTM models, with MWMAE-Large achieving the best overall performance across all evaluated audio representations.
While SSAMs outperform AxLSTMs slightly in overall score $s(m)$, there are more nuances in their performance characteristics on specific types of audio tasks. SSAM models perform extremely well in Speech-based and Audio classification tasks, especially for emotion recognition (CD) and spoken language identification (VL), whereas AxLSTMs fare better for music and pitch perception tasks (BO, Mri-T, NSynth-5h). While neither Mamba-based SSAMs nor xLSTM-based AxLSTMs outperform top of the line MAE-style MSMs, they consistently outperform directly comparable standard Transformer based counterparts, highlighting their viability in learning self-supervised audio representations.

\subsection{Ablations}

In this section, we compare Transformer, Mamba and xLSTM based MSMs and see how they fare as we manipulate two key hyperparameters: patch size and input sequence length during pretraining. For more ablations related to selection of other hyperparameters, refer to \cite{yadav2024masked, yadav2025axlstms}.
\subsubsection*{Patch size} As mentioned in Section~\ref{ssec:patching}, patch size is a key hyperparameter. Not only does it impact the length of the input, it also impacts the time-frequency resolution of the patches that are processed by the MSM, and thus, impacting model characteristics. The objective of this ablation is not only to see which models maintain performance with different patch sizes, thus indicating better capability to model distinct time-frequency resolutions, but also to observe how distinct sequence lengths impact performance. To this end, we pretrain and compare SSAST, SSAM and AxLSTM tiny configurations with 3 different patch configurations: (i) $f=16, t=8$, (ii), $f=16, t=4$ (default) and (iii) $f=8,t=4$, where conditions (i) and (iii) represent worsening frequency resolution and improving frequency resolution compared to the default setting (ii), respectively.
\begin{table}[t]
    \caption{Patch size ablations with the Tiny configuration}
    \small
    \label{tab:patchablations}
    \setlength\tabcolsep{5pt}
    \centering
    \begin{tabular}{l|ccc}
        \toprule
        Model & Patch parameters & \# Patches & $s(m)$\\
        \midrule
        SSAST-Tiny & $f=16, t=8$ & 125 & 46.8\ci{0.3} \\
        SSAM-Tiny & $f=16, t=8$ & 125 & \textbf{57.4\ci{0.3}}\\
        AxLSTM-Tiny &  $f=16, t=8$ & 125 & 55.1\ci{0.2} \\
        \midrule
        SSAST-Tiny & $f=16, t=4$ & 250 & 55.3\ci{0.2} \\
        SSAM-Tiny & $f=16, t=4$ & 250 & \textbf{75.0\ci{0.2}}\\
        AxLSTM-Tiny &  $f=16, t=4$ & 250 & 70.1\ci{0.2} \\
        \midrule
        SSAST-Tiny & $f=8,t=4$ & 500 & 53.5\ci{0.3} \\
        SSAM-Tiny & $f=8,t=4$ & 500 & \textbf{74.8\ci{0.2}}\\
        AxLSTM-Tiny & $f=8,t=4$ & 500 & 72.2\ci{0.3} \\
        \bottomrule
    \end{tabular}
\end{table}
From Table~\ref{tab:patchablations}, we can see that SSAMs and AxLSTMs outperform standard Transformer based SSAST across all patch settings. It is worth noting that while decreasing time resolution ($f=16, t=8$) worsens performance for all the models, SSAMs observe the biggest performance drop when time resolution is worsened. On the other hand increasing frequency resolution $f=8,t=4$ only helps the overall score of the AxLSTM-Tiny model.

\subsubsection*{Input sequence length during pretraining} In an attempt to tie up the impact of sequence length on downstream performance, we conduct a final ablation experiment where we keep the default patch configuration ($f=16, t=4$) while increasing only the duration of audio clip at the time of pretraining. AudioSet has clips of up to 10 seconds in duration, and by default we select a 2-second random crop during pretraining. So we conduct experiments for two additional settings: randomly clipping $5$ and $10$ second audio clips.
\begin{table}[t]
    \caption{Comparing SSAST, SSAM and AxLSTM overall performance with varying audio clip durations}
    \small
    \label{tab:clipduration}
    \setlength\tabcolsep{5pt}
    \centering
    \begin{tabular}{l|ccc}
        \toprule
        Model & duration (seconds) & \# Patches & $s(m)$\\
        \midrule
        SSAST-Tiny & $2$ & 250 & 55.3\ci{0.2} \\
        SSAM-Tiny & $2$ & 250 & \textbf{75.0\ci{0.2}}\\
        AxLSTM-Tiny &  $2$ & 250 & 70.1\ci{0.2} \\
        \midrule
        SSAST-Tiny & $5$ & 625 & 53.4\ci{0.3} \\
        SSAM-Tiny & $5$ & 625 & \textbf{75.4\ci{0.2}} \\
        AxLSTM-Tiny &  $5$ & 625 & 70.3\ci{0.4} \\
        \midrule
        SSAST-Tiny & $10$ & 1250 & 46.4\ci{0.4} \\
        SSAM-Tiny & $10$ & 1250 & \textbf{69.6\ci{0.2}} \\
        AxLSTM-Tiny & $10$ & 1250 & OOM \\
        \bottomrule
    \end{tabular}
\end{table}
From Table~\ref{tab:clipduration}, we can see that model performance improves for SSAM-Tiny when going from 2-second audio clips to 5-second audio clips, whereas there is very marginal improvement in performance for AxLSTMs. On the other hand, increasing clip duration consistently worsens performance for SSAST-Tiny models. Increasing clip duration any further worsens performance drastically for SSAMs, potentially indicating that we might have hit the performance limit for this model size. AxLSTM-Tiny threw an Out of Memory (OOM) error when attempting to train on 10-second inputs. 

OOM issues not withstanding, results from these two ablations indicate that SSAMs and AxLSTMs scale better with increasing sequence length compared to SSASTs, and are also more parameter efficient.

\section{Things we could not cover}
\label{sec:closing}

In this work, we discuss masked spectrogram modeling and recent advances in neural sequence modeling in context of learning self-supervised general-purpose audio representations. We discuss the two most prominent frameworks for masked spectrogram modeling, viz. SSAST-style and MAE-style MSMs, as well as dwelling into selective state space models (Mamba) and extended long short-term memory models (xLSTMs), which we put to the test against the Transformer architecture in a unified, reproducible MSM framework. However, we did not dive deeper into throughput and efficiency characteristics of the models. Efficiency and throughput is very hardware dependent, with certain implementations and kernels working better on certain hardware versus others. Testing all available implementations on a multitude of hardware devices in a multitude of sequence length settings is beyond the scope of this paper, whereas testing in a restricted scope have already been done for both Mamba \cite{gu2023mamba, mambav2, shams2024ssamba} and xLSTM \cite{beck2025xlstm}. Furthermore, there are too many nuances to keep track of for effective throughput analysis: things such as the interplay of JIT or runtime compilation, fused implementations, and even running on shared hardware in a university cluster setting, can obfuscate throughput analysis. We thus refrained from conducting a thorough throughput analysis in the paper.

Further, our analysis excludes experiments on full supervised finetuning of the models. While it might have been useful for certain readers and applications, running full finetuning experiments on all these models across all the datasets evaluated was simply infeasible. Besides, we firmly believe that the conducted experiments serve as a solid proxy for the expected level of performance on similar/adjacent tasks, and with our accompanying code and pretrained models, the interested readers would be able to conduct those experiments themselves.

\section{Conclusions}
\label{sec:conclusion}

In this work, we have presented a systematic overview of two key topics: (i) masked spectrogram modeling for learning self-supervised audio representation, and (ii) recent advances in neural sequence modeling architectures, viz. Mamba and xLSTM. The core objective of the paper is to present the reader a means to navigate the research landscape surrounding these topics as well as at the intersection of these topics, while laying down a solid foundation of the key underlying concepts at the heart of these topics. We provide a succinct yet thorough description of masked spectrogram modeling and the two main classes of masked spectrogram models (MSMs). Further, we provide a solid coverage of the fundamentals behind Mamba and xLSTM, the two most promising recent advances in neural sequence processing. We also provide an evaluation of Transformer, Mamba and xLSTMs for learning self-supervised audio representations through masked spectrogram modeling in a unified and reproducible framework, where through empirical analysis of ten varied downstream tasks, we establish the effectiveness of Mamba and xLSTM based approaches over a standard Transformer. All this was accompanied by a continued discussion about potential avenues for future research, and we hope that we inspired new research in the field.

\section*{Acknowledgments}
We would like to thank the Pioneer Centre for AI for supporting this research.

\bibliographystyle{IEEEtran}
\bibliography{IEEEabrv, references}


 




\vfill

\end{document}